\documentclass[conference,a4paper]{IEEEtran}
\IEEEoverridecommandlockouts

\usepackage[hidelinks]{hyperref}
\usepackage[cmex10]{amsmath}
\usepackage{amssymb,amsfonts}
\usepackage{dblfloatfix}

\usepackage[ruled,vlined]{algorithm2e}
\usepackage{graphicx}
\graphicspath{{Figures/PDF/}{Figures/PNG/}}

\usepackage{booktabs}
\usepackage{siunitx}
\usepackage[numbers,compress]{natbib}
\usepackage{texnames}
\usepackage{bm,bbm}
\usepackage{orcidlink}
\usepackage{pifont}
\usepackage{multirow}
\usepackage{makecell}

\begin{document}
\title{\uppercase{Radar Intelligent Detection with Coarse-Grained Labels}
\thanks{Corresponding author: Yumiao Wang (wangyumiao@uestc.edu.cn). \\
	\indent Thanks to the National Natural Science Foundation of China under Grant 62271126 and Grant U24B20188 for funding. 
	}
}

\author{\IEEEauthorblockN{Chuanfei Zang\orcidlink{0009-0005-8851-5926}, Yumiao Wang\orcidlink{0000-0002-1474-0486}, Xingyu Chen\orcidlink{0009-0001-4295-9928}, Yanbin Wang\orcidlink{0009-0007-0121-1095}, Guolong Cui\orcidlink{0000-0001-5707-6311}, Xiaobo Yang\orcidlink{0009-0000-4888-4050
	}}
	\IEEEauthorblockA{\textit{University of Electronic Science and Technology of China}, 611731 Chengdu, China
	}
}

\maketitle
\begin{abstract}
	Many deep learning based radar target detectors rely on range-cell level labels for training, which are expensive to obtain. To reduce the labeling burden, this paper presents a training strategy that uses only range-window level labels. Specifically, two sub-echoes are randomly cropped from the same window echo, resulting in a known relative shift between them. A siamese Transformer is then used to identify target-salient responses from one sub-echo and form pseudo labels, which are mapped to the paired sub-echo according to the known shift. In addition, a shift-equivariant consistency (SEC) constraint is imposed to align the predicted responses of the two sub-echoes after shift compensation. By exploiting shift priors within each labeled window, the proposed approach learns reliable detection without requiring cell-level labels. Experimental results show that it consistently surpasses classical constant false alarm rate (CFAR) detectors and achieves performance close to range-cell level labels training, while substantially reducing labeling cost.
\end{abstract}
\begin{IEEEkeywords}
	Radar intelligent detection, range-window labels, deep learning.
\end{IEEEkeywords}

\section{Introduction}
Radar, with its long detection range and robustness against weather and lighting variations, remains an irreplaceable sensor for target detection in search and rescue, surveillance, and national defense \cite{8767932}. Statistical detectors usually assume a specific clutter distribution; when this assumption is violated in complex environments, their performance may degrade noticeably \cite{4276955}. Feature-based methods design discriminative descriptors from Doppler, time-frequency, or phase-related information, but their generalization across diverse clutter conditions and target types is often limited \cite{5200356}.

To improve adaptability in complex environments, deep learning based approaches have been increasingly explored for radar detection because they can learn discriminative representations directly from data and capture complicated signal patterns \cite{10718342,10588477}. Representative models include convolutional neural networks (CNNs), graph neural networks, and long short-term memory networks (LSTMs), which have been adopted to build intelligent detectors and have shown improved performance in various scenarios \cite{11124398, 9990130, 11243251}. For instance, Qu et al. introduced an attention-enhanced CNN to extract deep features from time-frequency representations of radar signals \cite{9336010}. Xu et al. \cite{10582311} proposed a detector based on a spatiotemporal graph attention network to model inherent relationships in radar echoes. Wan et al. \cite{9854805} combined time-frequency features with a bidirectional LSTM for sequence-based detection.

Nevertheless, these detectors are trained using echo data with range-cell level labels, where each range cell is explicitly labeled as target or clutter to indicate the exact target location \cite{10848402,10210300}. In practical engineering, obtaining such fine-grained labels is expensive and often requires extensive field trials, reliable reference sensors, and stringent labeling procedures \cite{10361558}. In contrast, coarse-grained range-window level labels, which only indicate whether a target exists within a given range window, are much easier to collect in many applications. The key difficulty is that these coarse labels do not directly specify where the target lies inside the window, thereby making it particularly difficult to train a detector that can produce accurate range-cell level responses. To the best of our knowledge, this problem has not been systematically investigated in the radar detection literature, and this paper provides an early attempt to bridge this gap.

\begin{figure*}[t]
	\centering
	\includegraphics[width=0.90\linewidth, height=0.46\textwidth]{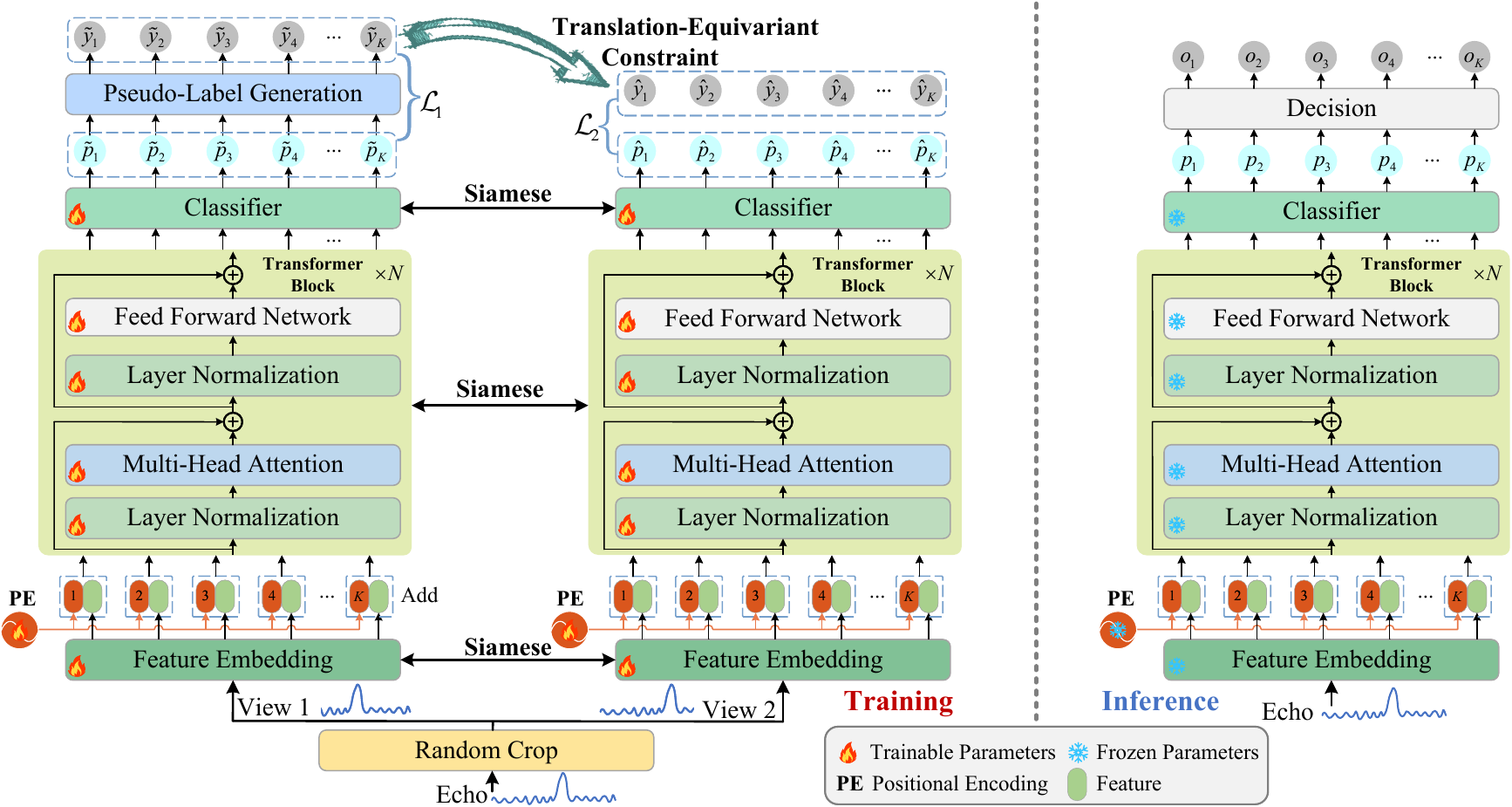}
	\vspace{-0.3cm}
	\caption{Model training and inference process of the proposed method.}
	\label{fig:model}
	\vspace{-0.7cm}
\end{figure*}

To address this issue, we propose a training method based on the siamese Transformer and the SEC constraint, as illustrated in Fig.~\ref{fig:model}. The proposed method uses only range-window level labels during training while achieving competitive detection performance. Our work makes the following contributions:
(1) We propose a siamese Transformer architecture with parameter sharing to enable collaborative learning from paired views, improving feature robustness and sensitivity to target-related responses.
(2) We introduce an SEC constraint that explicitly incorporates the shift-equivariance prior along the range dimension. This constraint enforces prediction consistency between the paired views, guiding the model to infer precise target locations from range-window level labels.
(3) We conduct systematic experiments on measured radar datasets to validate the effectiveness of the proposed method.

\section{Methodology}
\subsection{Signal Model}
Consider a radar transmitting a linear frequency modulated waveform for target detection:
\begin{equation}
	s(t) = {\rm{rect}}\left( {{t}/{T_{\text{r}}}} \right)\exp \left( j2\pi {{f_{\text{c}}}t + j\pi k{t^2}} \right),
\end{equation}
where $f_{\text{c}}$ is the carrier frequency, $k$ is the chirp rate, and $T_{\text{r}}$ is the pulse width. For a point target at range $R$, the received signal is
\begin{equation}
	r(t) = \alpha s(t - {\tau}) + c(t) + n(t),
\end{equation}
where $\tau = 2R/c_0$, $c_0$ is the speed of light, $\alpha$ is the complex coefficient accounting for target reflectivity and propagation attenuation, $c(t)$ represents clutter, and $n(t)$ denotes additive white Gaussian noise. After standard signal processing, we obtain the baseband echo $\bm{r} \in \mathbb{C}^{M \times 1}$, where $M$ is the number of range cells. The target is assumed to lie at the $m_0$-th range cell with $m_0=\lceil \tau f_{\text{s}} \rceil$, where $f_{\text{s}}$ is the sampling rate and $\lceil \cdot \rceil$ denotes the ceiling function.

Conventional intelligent detectors take $\bm{r}$ as input and output the posterior probability of target presence in each range cell. In this work, only range-window level labels $y\!\in\!\{0,1\}$ are provided: $y=1$ indicates that the window contains a target, whereas $y=0$ indicates no target. Since this label lacks precise target position information within the window, directly training the model with the entire echo and the label $y$ would prevent the model from detecting the specific range cells where targets are located. To mitigate this issue, we construct two sub-echoes from the same labeled window with a known relative shift, and use this shift information to tie their predicted responses of the models to a physically consistent target location. This is implemented with a siamese Transformer with shared parameters and SEC constraint.

Specifically, consider a labeled range-window echo of length $S$ with label $y$. To incorporate more clutter information, we extend it by $L$ range cells on both sides and obtain an extended echo ${\bm x}\in \mathbb{C}^{(S+2L)\times 1}$. We assume that each window contains at most one target and that the extended $L$ cells do not contain a target. Two sub-echoes of length $K$, denoted by $\tilde{\bm x}$ and $\hat{\bm x}$, are then randomly cropped from ${\bm x}$, and we set $S=K-L$ so that each cropped sample of length $K$ includes the same labeled range-window of length $S$. Therefore, $\tilde{\bm x}$ and $\hat{\bm x}$ inherit the same label $y$, and they can be regarded as two views of the same potential target. The relative shift between $\tilde{\bm x}$ and $\hat{\bm x}$ is determined by their cropping offsets and is thus known during training. Furthermore, by stacking the real and imaginary parts of each sub-echo as two channels, we form ${\bm{\tilde X} },{\bm {\hat X}}\in\mathbb{R}^{2\times K}$, which are fed into the dual branches of the Transformer. The two branches share network parameters and are jointly optimized with the SEC constraint, which enforces agreement between the two predicted response maps after compensating for the known shift. In this way, the model can be trained using window-level labels while still learning accurate range-cell level target responses.

\subsection{Siamese Transformer}\label{Siamese}
The proposed siamese Transformer consists of two branches with identical architectures and shared parameters. Each branch sequentially comprises a feature embedding module, a positional encoding module, $N$ Transformer blocks, and a classifier, as shown in Fig.~\ref{fig:model}. For clarity, we describe one branch with input ${\bm X}$. A single convolutional layer is first applied for feature embedding, producing a feature representation ${\bm F}\in{\mathbb R}^{C\times K}$, where $C$ is the number of feature channels.

Since the Transformer block is inherently insensitive to the absolute positions of input features, while positional information along the range dimension is crucial for radar target detection. Therefore, we introduce a positional encoding module to inject range-dependent information. Specifically, the positional sequence ${\bm h}=[1,2,\dots,K]$ is mapped to a positional encoding ${\bm P}\in\mathbb{R}^{C\times K}$ via a fully connected (FC) layer, and the result is added to ${\bm F}$ to obtain the position-aware input:
\begin{equation}
	{\bm E}_0 = {\bm F} + {\bm P},\quad{\bm F} = {\text {conv}}( {\bm X}),\quad {\bm P}={\text {FC}}\left( \bm h\right).
\end{equation}
The Transformer blocks then take ${\bm E}_{i-1}$ as input to extract more discriminative features. Each block follows a standard pre-normalization structure with layer normalization (Norm)~\cite{ba2016layer}, multi-head attention (MHA)~\cite{vaswani2017attention}, and a feed-forward network (FFN)~\cite{touvron2023llama}, together with residual connections:
\begin{equation}
	\begin{aligned}
		\bar{\bm E}_{i-1} = {\bm E}_{i-1} + {\text {MHA}}\left( {\text {Norm}}\left({\bm E}_{i-1}\right) \right),
	\end{aligned}
\end{equation}
\begin{equation}
	\begin{aligned}
		{\bm E}_i = \bar{\bm E}_{i-1} + {\text {FFN}}\left( {\text {Norm}}\left(\bar{\bm E}_{i-1}\right) \right) .
	\end{aligned}
\end{equation}
Here, Norm normalizes features to enhance training stability and mitigate the impact of feature amplitude fluctuations caused by clutter on model convergence and detection decisions. MHA integrates feature information across all range cells to model the target's response pattern over neighboring cells and its relationship with surrounding clutter. FFN enhances the separability between targets and clutter in the feature space through nonlinear transformations.

Finally, the classifier produces a range-cell-wise decision score ${\bm z}\in{\mathbb R}^{K \times 1}$ from ${\bm E}_{N}$ and converts it to a posterior probability ${\bm p}\in{\mathbb R}^{K \times 1}$ using a sigmoid function:
\begin{equation}
	{\bm z} = {\text {FC}}\left( {\bm E}_{N}\right),\quad
	{\bm p} = {1}/\left( {1 + \exp\left( {-{\bm z}}\right) }\right) .
\end{equation}
The element ${\bm p}_k$ denotes the posterior probability that the $k$-th range cell contains a target. For the two echo views, the resulting posterior probability vectors are denoted by ${\bm {\tilde p}}$ and ${\bm{\hat p}}$, respectively.

\subsection{Shift-Equivariant Consistency Constraint}
Given the posterior probability vector of view~1, $\tilde{\bm p}$, we adaptively construct a range-cell level pseudo label ${\tilde{\bm y}}\in{\mathbb R}^{K\times 1}$ according to the range-window label $y$. When $y=0$, we set $\tilde{\bm y}=\bm 0$. When $y=1$, considering that each range window contains at most one target, we treat the range cell with the maximum posterior probability in $\tilde{\bm p}$ as the target location and the remaining cells as non-target:
\begin{equation}
	k^\star = \arg\max_{1 \le k \le K} \tilde{\bm p}_k, \quad
	\tilde{\bm y}_k =
	\begin{cases}
		1, & k = k^\star,\\
		0, & \text{otherwise}.
	\end{cases}
\end{equation}

We then exploit the known cropping-induced shift between the two sub-echoes to construct a shift-aligned pseudo label for view~2. Let $\Delta$ denote the relative cropping offset (in range cells) from view~1 to view~2, and define a shift operator $S_\Delta(\cdot)$ as
\begin{equation}
	[S_\Delta(\tilde{\bm y})]_k =
	\begin{cases}
		{\tilde{\bm y}}_{k+\Delta}, & 1 \le k+\Delta \le K,\\
		0, & \text{otherwise}.
	\end{cases}
\end{equation}
The pseudo label for view~2 is then given by $\hat{\bm y}=S_\Delta(\tilde{\bm y})$. This construction explicitly injects the prior knowledge of shift-equivariance along the range dimension: when the input echo is shifted in range, the predicted response should shift by the same amount, thereby ensuring physical consistency between the two views.

With the above pseudo labels, the training objective is defined as the equally weighted average of two loss terms:
\begin{equation}
	{\cal L} = {\left( {\cal L}_1 + {\cal L}_2\right) }/{2}, \quad
	{\cal L}_1 = \ell(\tilde{\bm p}, \tilde{\bm y}), \quad
	{\cal L}_2 = \ell(\hat{\bm p}, \hat{\bm y}),
\end{equation}
where $\ell(\cdot,\cdot)$ is the per-range-cell binary cross-entropy loss:
\begin{equation}
	\ell(\tilde{\bm p}, \tilde{\bm y})
	= - \frac{1}{K} \sum_{k=1}^{K} \big[ \tilde{\bm y}_k \log \tilde{\bm p}_k + (1 - \tilde{\bm y}_k)\log(1 - \tilde{\bm p}_k) \big].
\end{equation}
By requiring the predictions from both views to agree with their shift-aligned pseudo labels, the model is encouraged to produce sharp, range-resolved target responses while maintaining cross-view consistency. This constraint improves the reliability of the learned response patterns and, in turn, enhances target detection performance.

\subsection{Inference}
During inference, only a single branch of the siamese Transformer is retained as the target detector. Unlike the random cropping strategy used in training, the input echo is partitioned along the range dimension using a sliding window of length $K$ at test time. Each windowed sample is fed into the model sequentially to obtain the range-cell-wise posterior probability vector ${\bm p}$ for the corresponding window. A threshold $h=0.5$ is then applied to make decisions for each range cell, yielding the binary detection output vector ${\bm o}$ using the indicator function $\mathbb{I}(\cdot)$:
\begin{equation}
	{\bm o}_k = \mathbb{I}\big( {\bm p}_k > h \big), \quad k = 1,\dots,K.
\end{equation}

\section{Experiments}
\begin{table}[t]
	\renewcommand{\arraystretch}{0.2}
	\caption{Radar Parameters of MTDSP Database}
	\vspace{-0.3cm}
	\label{Radar_parameters}
	\centering
	\setlength{\tabcolsep}{2.0mm}{ 
		\vspace{-0.1cm}
		\begin{tabular}{cccc}
			\bottomrule
			\cmidrule{1-4}  
			{\textbf{Band Width}} & {\textbf{Carrier Frequency}} & {\textbf{PRF}} & {\textbf{Sample Frequency}} \\ 
			\cmidrule{1-4} 
			25 MHz & 9.4 GHz & 2000 Hz & 60 MHz\\
			\bottomrule
			\cmidrule{1-4}
	\end{tabular}}
	\vspace{-0.7cm}
\end{table}

\subsection{Experimental Setup}
{\bf{Dataset Description:}} Experiments are conducted on the HH-polarization echo file 20221114050022\_stare\_HH.mat from the marine target detection and sharing program~\cite{jian2023sea}. The data were collected in staring mode by an SPPR50P X-band solid-state coherent radar deployed in Yantai, Shandong, China, on November 12, 2022. In total, 131{,}072 pulses were recorded, and the radar system parameters are summarized in Table~\ref{Radar_parameters}. To evaluate detection performance under different signal-to-clutter ratios (SCR), we adopt a hybrid data generation strategy: point target echoes are simulated using the radar system parameters and then superimposed onto real sea clutter backgrounds. This strategy preserves the statistical characteristics of measured sea clutter while enabling precise control of the target strength. To avoid potential interference from extended targets present in the raw measurements, range cells 130--450 and 680--1000 are selected as clutter-only backgrounds. Simulated targets are superimposed at range cells 300 and 850, respectively.

{\bf{Implementation Details:}} Following the data construction procedure in Section \ref{Siamese}, the window extension length is set to $L=8$, and the cropped echo length is set to $K=64$. The constructed dataset is split into training, validation, and test sets with a ratio of 6:2:2. For the proposed model, the number of Transformer blocks is set to $N=6$. The model is trained with the AdamW optimizer~\cite{10480574} for 200 epochs using a batch size of 256. All experiments are performed on a server equipped with 8 NVIDIA A40 GPUs. During evaluation, a detection is counted as correct if the predicted target range cell matches the ground-truth cell or falls within one adjacent cell on either side.

\subsection{Effectiveness Evaluation}
To validate the effectiveness of the proposed method and analyze the contribution of each component, we conduct ablation experiments on the dataset with SCR=10~dB and summarize the results in Table~\ref{Ablation}. Specifically, “\ding{55} Siamese” removes the siamese architecture: during training, only the view~1 branch is involved in backpropagation, and its gradients are used to update the shared parameters. “\ding{55} Transformer” replaces the Transformer backbone with CV-UNet~\cite{9919196} while keeping the rest of the training procedure unchanged.

The ablation results indicate that both the siamese architecture and the SEC constraint are essential for learning target detection. Removing either component prevents the model from converging to an effective detector. Moreover, when the siamese architecture and the SEC constraint are both retained, replacing the Transformer network with CV-UNet yields comparable detection performance. This observation suggests that the proposed training strategy is largely model-agnostic: its effectiveness mainly comes modeling and exploiting the shift-equivariance physical prior of radar echoes, rather than relying on a carefully tailored backbone.

\begin{table}[t]
	\vspace{-0.4cm}
	\renewcommand{\arraystretch}{0.4}
	\caption{Impact of Key Components on the Performance of Model}
	\label{Ablation}
	\centering
	\setlength{\tabcolsep}{1.7mm}  
	\vspace{-0.4cm}
	\begin{tabular}{c|ccc||cc}
		\bottomrule
		\cmidrule{1-6}
		& \textbf{Transformer} & \textbf{Siamese} & \textbf{SEC} & $P_{\text{d}}$ & $P_{\text{fa}}$  \\  
		\bottomrule
		\cmidrule{1-6} 
		\multirow{4}{*}{\textbf{Method}} & \ding{51}  & \ding{51}  & \ding{55} & 0.35\% & $3.8\times10^{-3}$\\
		& \ding{51}  & \ding{55}  & \ding{51}& 0.12\%  & $2.6\times10^{-3}$\\ 
		& \ding{55}  & \ding{51}  & \ding{51} & 93.75\% & $1.1\times10^{-6}$\\
		& \ding{51}  & \ding{51}  & \ding{51} & {\bf{99.99\%}} & $\bm{5.6\times10^{-7}}$\\
		\bottomrule
		\cmidrule{1-6}
	\end{tabular}
	{\footnotesize $\bullet$ \ding{51} and \ding{55} indicate use or not use of this approach.}
	\vspace{-0.4cm}
\end{table}

\begin{figure}[t]
	\centering
	\includegraphics[width=1.0\linewidth, height=0.37\textwidth]{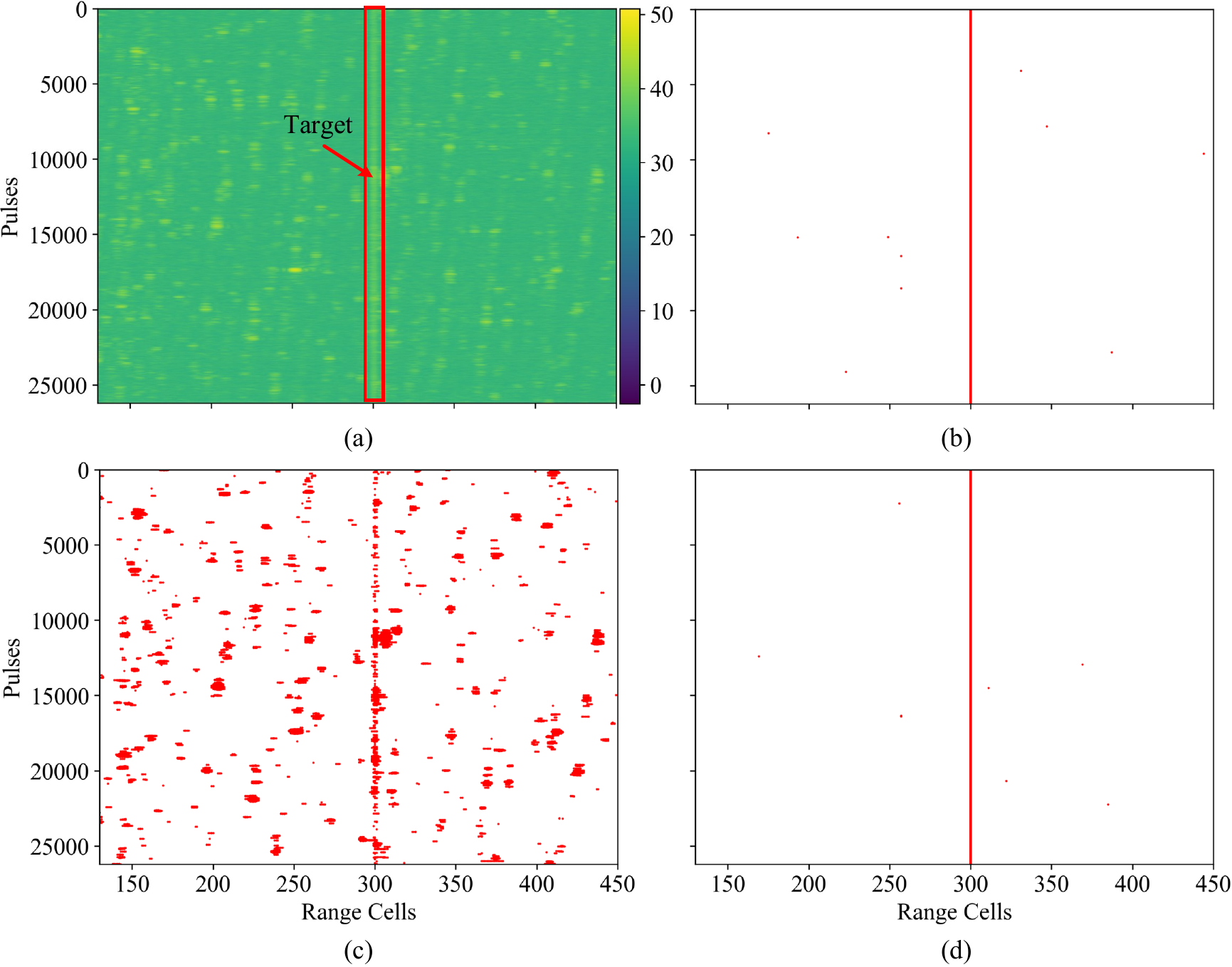}
	\vspace{-0.8cm}
	\caption{Detection result. (a) Radar echoes; (b) Proposed method; (c) OS-CFAR; (d) Performance boundary;}
	\label{detect_result}
	\vspace{-0.7cm}
\end{figure}
\subsection{Performance Comparison}
To illustrate the detection performance of the proposed method under complex sea clutter backgrounds, we compare our detector with two representative detectors at SCR=5~dB: (i) a reference detector trained with fine-grained range-cell level labels following the setting in~\cite{9919196}, and (ii) a traditional OS-CFAR detector. The reference detector in (i) adopts the same network architecture and inference pipeline as our detector, and differs only in the annotation detail used for training.  Therefore, it can be regarded as performance boundary for quantifying, under an identical model and inference procedure, how much performance is affected when replacing fine-grained range-cell level labels with coarse range-window level labels. The input echoes and the corresponding detection results are shown in Fig.~\ref{detect_result}. As can be seen, under strong clutter, OS-CFAR exhibits noticeable false alarms and missed detections. In contrast, the proposed method correctly detects the target while maintaining a false-alarm level comparable to that of the reference detector in (i).

\begin{figure}[t]
	\centering
	\includegraphics[width=1.0\linewidth, height=0.20\textwidth]{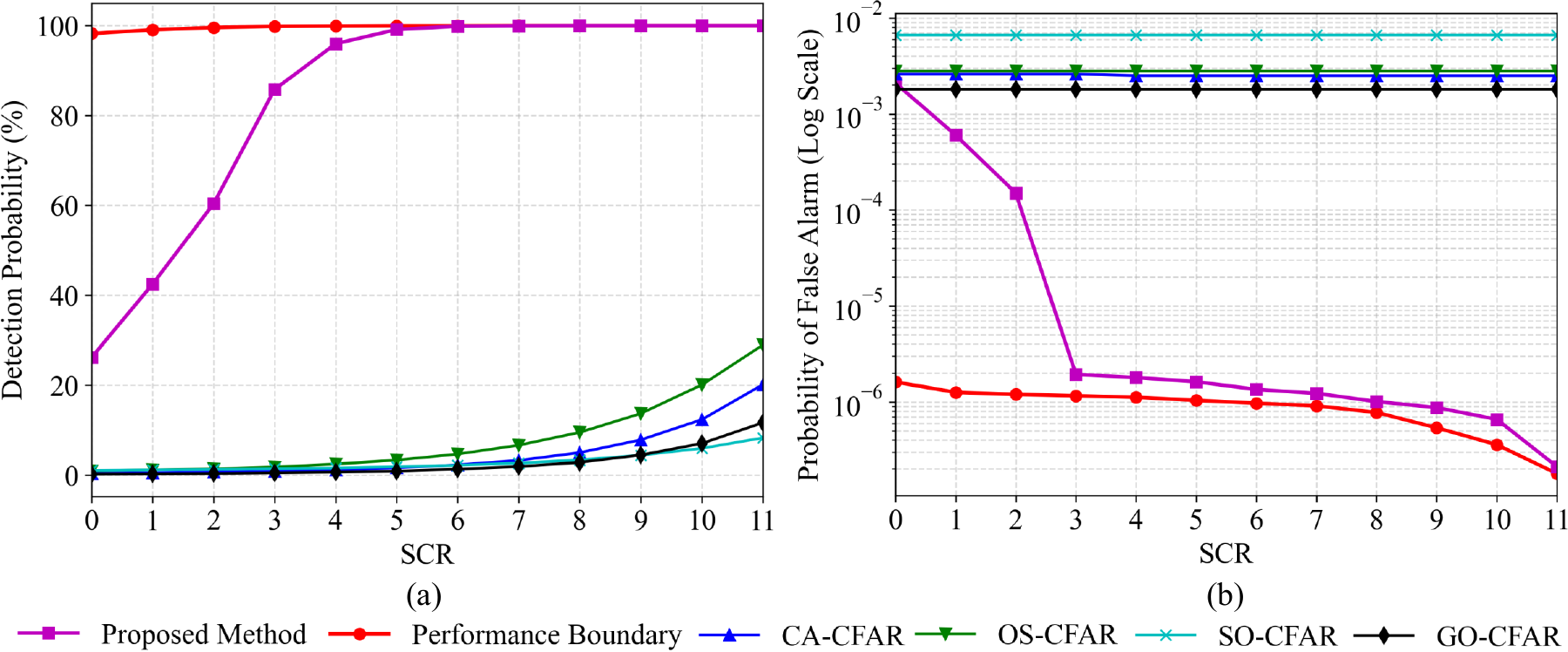}
	\vspace{-0.8cm}
	\caption{Comparison of detection performance among different methods. (a) Probability of detection; (b) Probability of false alarm.}
	\label{pd_pfa}
	\vspace{-0.7cm}
\end{figure}

To further quantify performance over a wider SCR range, we vary the target SCR from 0 to 11~dB, and extend the CFAR baselines to CA-CFAR, OS-CFAR, SO-CFAR, and GO-CFAR. The corresponding $P_{\text{d}}$ and $P_{\text{fa}}$ curves are reported in Fig.~\ref{pd_pfa}. The results show that the proposed method consistently achieves a higher detection probability than all CFAR variants across the entire SCR range while keeping a low false-alarm rate. Compared with reference detector (i) trained with range-cell level labels, our method still exhibits a performance gap in the low-SCR regime of 0--2~dB; however, when SCR reaches 3~dB and above, the performances of the two detectors become closely matched. 

Overall, the above results suggest that using only range-window level labels during training can achieve detection performance close to that obtained with range-cell level labels in most SCR conditions. This substantially reduces the reliance on costly fine-grained labeling with virtually no loss in detection performance, which in turn lowers data preparation effort and facilitates practical deployment of intelligent radar detectors.

\section{Conclusion}
To address the practical difficulty of obtaining fine-grained range-cell level labels in engineering applications, this paper presents a radar target detection training method that requires only coarse range-window level labels. The proposed approach adopts a siamese Transformer architecture, where echoes within the same range window are cropped into two different views and collaboratively learned through parameter sharing, thereby enhancing feature representation and improving sensitivity to target location variations. In addition, we introduce a shift-equivariant consistency constraint that explicitly incorporates the range-dimension shift prior into the training process, enabling the model to learn range-cell-level target detection responses even under window-level supervision. Experimental results demonstrate that the proposed method significantly outperforms classical CFAR detectors and achieves performance comparable to fully supervised methods with fine-grained labels under most SCR conditions, thus substantially reducing annotation difficulty and data preparation cost with little sacrifice in detection performance.

\small
\bibliographystyle{IEEEtranN}
\bibliography{references}

\end{document}